\documentclass[twocolumn,prb,superscriptaddress,noshowpacs,epsf]{revtex4}

\usepackage{graphicx}

\begin{document}
\title{Persistent single-photon production by tunable on-chip micromaser
with a superconducting quantum circuit}
\date{\today}
\author{J. Q. You}
\affiliation{Department of Physics and Surface Physics Laboratory (National Key
Laboratory), Fudan University, Shanghai 200433, China}
\affiliation{Frontier Research System, The Institute of Physical
and Chemical Research (RIKEN), Wako-shi 351-0198, Japan}
\author{Yu-xi Liu}
\affiliation{Frontier Research System, The Institute of Physical
and Chemical Research (RIKEN), Wako-shi 351-0198, Japan}

\author{C. P. Sun}
\affiliation{Frontier Research System, The Institute of Physical
and Chemical Research (RIKEN), Wako-shi 351-0198, Japan}
\affiliation{Institute of Theoretical Physics, Chinese Academy of Sciences,
Beijing 100080, China}

%
\author{Franco Nori}
\affiliation{Frontier Research System, The Institute of Physical
and Chemical Research (RIKEN), Wako-shi 351-0198, Japan}
\affiliation{Center for Theoretical Physics, Physics Department,
Center for the Study of Complex Systems,
University of
Michigan, Ann Arbor, MI 48109-1040, USA}

\begin{abstract}
We propose a tunable on-chip micromaser using a superconducting quantum circuit (SQC). 
By taking advantage of externally controllable state transitions,
a state population inversion can be achieved and preserved for the two working
levels of the SQC and, when needed, the SQC can generate a single
photon. We can regularly repeat these processes in each cycle
when the previously generated photon in the cavity is decaying,
so that a periodic sequence of single photons can be
produced persistently. This provides a controllable way for implementing a persistent
single-photon source on a microelectronic chip.
\end{abstract}
\pacs{85.25.-j, 42.50.Pq}
\maketitle

\section{Introduction}

Superconducting quantum circuits can behave like natural atoms
and are also promising candidates of qubits for scalable quantum
computing.\cite{YN05} Moreover, these circuits also show quantum optical
effects and provide exciting opportunities for demonstrating quantum effects
at macroscopic scales and for conducting atomic-physics experiments on a
microelectronic chip 
(see, e.g., Refs.~\onlinecite{YN05,CHIO,YN03,YANG,YALE04,LIU04,ZAGO,LIU05,NANO}).

Because of its fundamental importance in quantum communications, single-photon sources
are crucial in both quantum optics and quantum electronics.\cite{REVIEW}
Single-photon sources can be achieved using quantum-dot-based
devices (see, e.g., Ref.~\onlinecite{SPS}), but their frequencies are not 
in the microwave regime required for superconducting qubits. 
Recently, there have been efforts to generate single photons by coupling
a superconducting qubit to a superconducting resonator.\cite{YALE04,LIU04,MARI}
However, because of damping inside the resonator, the generated single photon
can only persist for a very short time.

Here we show how to persistently produce steady microwave single photons
by a {\it tunable} micromaser using a superconducting quantum circuit (SQC).
The physical mechanism is as follows: The SQC acts like a {\it controllable}
artificial atom (AA) and is placed in a quantum electrodynamic cavity.
By taking advantage of the
externally controllable state transitions, one can pump the AA
to produce state population inversion for the two working levels.
This population inversion is preserved by turning off the transition to the ground
state, but when needed this transition can be switched on to generate a photon.
Within the photon lifetime of the cavity, one can pump the superconducting
AA to produce the state population inversion again for the next cycle
of operations and then switch on the state transition when the
photon generated in the previous cycle is decaying. By periodically repeating this
cycle, one can generate single photons in a persistent way.

Steady-state photons can also be generated by a micromaser with
natural atoms (see, e.g., Refs.~\onlinecite{MASER} and \onlinecite{ORZ}). 
However, in such a micromaser, there is a very small number of excited atoms among
all atoms passing through the cavity and these excited atoms enter
the cavity at {\it random} times. This will produce large
fluctuations for the photon field of the cavity. For instance, an
excited atom can enter the cavity long before or after the
previously generated single photon decays. To overcome this
problem, a state population inversion is prepared for the
superconducting AA in each cycle and all the cycles are repeated
periodically. Also, the cavity can be realized using an on-chip
superconducting resonator 
so that both the SQC and the resonator
can be fabricated on a chip. This might be helpful for
transferring quantum information between superconducting qubits in
future applications. Moreover, in contrast to the fixed
difference between the two working energy levels in a natural
atom, the level difference for the superconducting AA is tunable,
providing flexibility for producing a single-photon source over a
wider frequency region. 

In Ref.~\onlinecite{BAR}, spontaneous and stimulated emission characteristics 
was investigated for a Josephson-junction-cavity system, 
but here we focus on the quantum electrodynamic effects in the strong-coupling  
regime for a superconducting AA in the on-chip cavity. 
Moreover, the circuit design in this approach provides an enhanced level of 
control that is desirable for producing a single-photon source.

\begin{figure}
 \includegraphics[width=3.4in,  
bbllx=69,bblly=383,bburx=516,bbury=607]{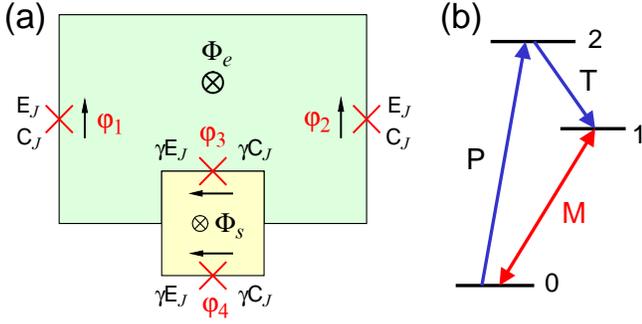} \caption{(Color
online) (a)~Schematic diagram of the superconducting artificial atom.
A symmetric SQUID and two identical Josephson junctions
with coupling energy $E_J$ and capacitance $C_J$ are placed in a
superconducting loop pierced by an externally applied magnetic
flux (green) $\Phi_e$. The two Josephson junctions in the
symmetric SQUID have coupling energy $\gamma E_J$ and capacitance
$\gamma C_J$, and the applied flux (yellow) threading through the
SQUID loop is $\Phi_s$. Here we choose $\gamma=0.5$ and
$E_J/E_c=100$, with $E_c=e^2/2C_J$ being the single-particle
charging energy of the junction. (b)~Transition diagram of a tunable 
artificial atom (i.e., a qutrit) used for the micromaser. 
A microwave field pumps the qutrit to make the transition 
$|0\rangle\longrightarrow |2\rangle$ (denoted as P) and another microwave field 
triggers the transition $|2\rangle\longrightarrow |1\rangle$ (denoted as T). 
With a state population inversion population inversion established for $|0\rangle$ and $|1\rangle$, 
the transition between $|0\rangle$ and $|1\rangle$ (denoted as M) is switched off 
to preserve the population inversion and, when needed, switched on to 
couple the AA with the cavity mode.} \label{fig1}
\end{figure}

\section{Superconducting artificial atom}

We consider an AA based on the SQC
for the flux qubit.\cite{FLUX}
 Instead, we use it as a qutrit 
involving the lowest three energy levels of the device.
Also, the fourth and other higher levels are well separated and not populated.
As shown in Fig.~1(a),
in addition to two identical Josephson junctions, a symmetric SQUID is placed
in the loop pierced by an external magnetic flux $\Phi_e$.
This SQUID increases the external controllability of the quantum circuit
by providing a tunable effective coupling energy: $\alpha E_J$
with $\alpha=2\gamma\cos(\pi\Phi_s/\Phi_0)$, where $\Phi_0$ is the flux quantum.

The Hamiltonian of the system is
\begin{equation}
H=\frac{P_p^2}{2M_p}+\frac{P_q^2}{2M_q}+U(\varphi_p,\varphi_q),
\end{equation}
with 
\begin{eqnarray}
P_i &\!=\!& -i\hbar\frac{\partial}{\partial\varphi_i},\;\;\;\;i=p,q,\nonumber\\
M_p &\!=\!& 2C_J (\Phi_0/2\pi)^2,\\ 
M_q &\!=\!& \frac{1}{4}M_p(1+4\gamma).\nonumber
\end{eqnarray}
The potential $U(\varphi_p,\varphi_q)$ is 
\begin{eqnarray}
U(\varphi_p,\varphi_q) &\!=\!& 2E_J[1-\cos\varphi_p\cos(\pi f+\frac{1}{2}\varphi_q)]
\nonumber\\
&&+2\gamma E_J[1-\cos(\pi f_s)\cos\varphi_q],
\end{eqnarray}
where $\varphi_p=(\varphi_1+\varphi_2)/2$ and $\varphi_q=(\varphi_3+\varphi_4)/2$.
The reduced fluxes $f_s$ and $f$ are given by
\begin{equation}
f_s=\frac{\Phi_s}{\Phi_0}\,,\;\;\;
f=\frac{\Phi_e}{\Phi_0}+\frac{f_s}{2}\,.
\end{equation}
The operator $P_k$ and the phase $\varphi_j$ 
obey $[\varphi_j,P_k]=i\hbar\delta_{jk}$, where $j,k=p,q$.

\begin{figure}
 \includegraphics[width=3.4in, 
bbllx=21,bblly=85,bburx=563,bbury=756]{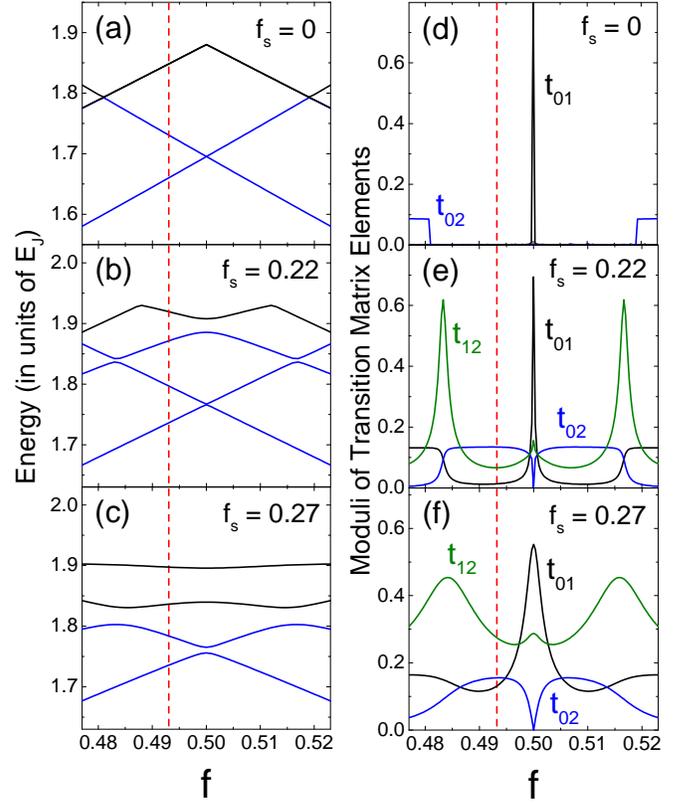}
\caption{(Color online) (a)-(c)~Energy levels of the superconducting artificial atom
versus the reduced magnetic flux $f=\Phi_e/\Phi_0+f_s/2$
for $f_s(\equiv \Phi_s/\Phi_0)=0$, $0.22$, and $0.27$, where only the lowest
four energy levels ($E_i$, $i=0$ to 3) are shown.
(d)-(f)~Moduli of the transition matrix elements $|t_{ij}|$
(in units of $I_c\Phi_w^{(0)}$) versus the reduced flux $f$ for $f_s=0$, $0.22$,
and $0.27$. Two vertical (red) dashed lines are plotted at $f=0.493$
as a guide to the eye.
}
\label{fig2}
\end{figure}

To make a transition between two energy levels $E_i$ and $E_j$ of
the superconducting AA, a microwave field
\begin{equation}
\Phi_w(t)=\Phi_w^{(0)}\cos(\omega_{ij}t+\theta)
\end{equation}
is applied through the larger superconducting loop of the quantum circuit.
For a weak microwave field, the time-dependent perturbation
Hamiltonian is 
\begin{equation}
H'(t)=-I\Phi_w(t),
\label{HA}
\end{equation}
and the transition matrix element between states $|E_i\rangle$ and
$|E_j\rangle$ is given by
\begin{equation}
t_{ij}=\langle E_i|\,I\,\Phi_w^{(0)}|E_j \rangle,
\end{equation}
where 
\begin{equation}
I=-I_c\cos\varphi_p\sin(\pi f+\frac{1}{2}\varphi_q)
\end{equation}
is the circulating supercurrent in
the loop without the applied microwave field, and the critical
current of the junction is defined as $I_c=2\pi E_J/\Phi_0$.

Figures 2(a)-2(c) display the dependence of the energy levels on the reduced flux $f$
for three different values of $f_s$. For a symmetric SQUID with $\gamma=0.5$,
these values of $f_s$ give rise to an effective Josephson coupling energy
$\alpha E_J$ with $\alpha=1$, $0.77$, and $0.66$, respectively.
At $f_s=0$, the third and fourth energy levels become degenerate and other
adjoining levels touch at the crossing points. When $f_s$ increases, this state
degeneracy is removed and gaps develop at the crossing points, more pronounced
for higher levels. In Figs. 2(d)-2(f),
we show the moduli of the transition matrix elements $|t_{ij}|$ for
the lowest three levels. At $f_s=0$, the transition matrix elements $t_{01}$,
$t_{02}$ and $t_{12}$ become zero in a wider region around $f=0.5$.
This means that the corresponding state transitions are forbidden.
With $f_s$ increasing, these state transitions become allowed, but the modulus
of each transition matrix element increases in a different manner.
In contrast, $|t_{01}|$ for the state transition between the two lowest levels
$E_0$ and $E_1$ increases slowly. Below we will explore these novel properties to
implement a micromaser using a quantum circuit on a chip.
Also, a superconducting ring containing only one Josephson junction can be used 
to achieve a qutrit (see, e.g., Ref.~\onlinecite{HANPRL}), 
but it requires a relatively large loop inductance, 
which makes the qutrit more susceptible to the magnetic-field noise.

\section{Fast adiabatic quantum-state control and state population inversion}

The state evolution of the superconducting AA depends on
the external parameters. For two given quantum states
$|E_i\rangle$ and $|E_j\rangle$, to have the evolution adiabatic,
the nonadiabatic coupling $\langle E_i|\frac{d}{dt}|E_j\rangle$ and the energy
difference $E_i-E_j$ should satisfy the condition (see, e.g., Ref.~\onlinecite{LIU05}):
\begin{equation}
\left|\frac{\hbar\langle E_i|(d/dt)|E_j\rangle}{E_i-E_j}\right|\ll 1.
\end{equation}
Here we change $f_s$ but keep the reduced flux $f$ unchanged. The adiabatic
condition can be rewritten as 
\begin{equation}
K_{ij}\left|\frac{df_s}{dt}\right|\ll 1, 
\end{equation}
where
\begin{equation}
K_{ij}=\left|\frac{\hbar\langle E_i|(\partial H/\partial f_s)|E_j\rangle}
{(E_i-E_j)^2}\right|.
\end{equation}

Figure~3 shows the quantities $K_{01}$ and $K_{12}$
as a function of the reduced flux $f$ for different values of $f_s$. For instance,
in the vicinity of $f=0.493$ (vertical dashed lines in Fig.~2),
$K_{01}\sim 0.2$ for $f_s=0.27$ [see Fig.~3(a)].
We can have 
$$K_{01}\left(\frac{df_s}{dt}\right)\sim 0.02 \ll 1$$ 
with $df_s/dt=0.1$~ns$^{-1}$, corresponding to a speed of changing $\Phi_s$,
$d\Phi_s/dt \sim 0.1\Phi_0$ per ns.
When $f_s=0.22$, 
$$K_{01}\left(\frac{df_s}{dt}\right)\sim 0.02$$ 
for $df_s/dt=2$~ns$^{-1}$,
and $d\Phi_s/dt$ can be much faster, to have the adiabatic
condition satisfied by decreasing $f_s$.
Also, around $f=0.493$, $K_{12}\sim 0.4$ for $f_s=0.27$ [see Fig.~3(b)].
When $df_s/dt=0.1$~ns$^{-1}$, 
$$K_{12}\left(\frac{df_s}{dt}\right) \sim 0.04\ll 1,$$ 
implying that the adiabatic condition is satisfied.
For a smaller $f_s$, $K_{12}$ decreases significantly and a much higher
$d\Phi_s/dt$ can be used.
This important property reveals that,
at $f\sim 0.493$, one can {\it adiabatically} manipulate the quantum states
$|0\rangle\equiv|E_0\rangle$ and $|1\rangle\equiv|E_1\rangle$ of the superconducting 
AA by quickly changing $f_s$ (e.g., $d\Phi_s/dt\agt 0.1\Phi_0$~ns$^{-1}$)
in the region of $0\leq f_s\alt 0.27$.

\begin{figure}
 \includegraphics[width=3.4in,  
bbllx=49,bblly=445,bburx=551,bbury=730]{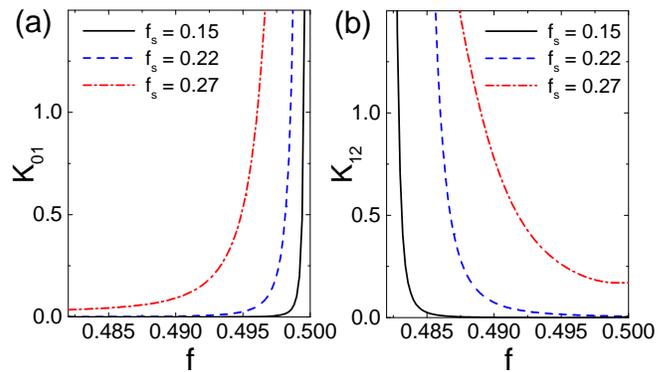}
\caption{(Color online) The quantities (a)~$K_{01}$ and (b)~$K_{12}$ for the adiabatic
condition versus the reduced flux $f$ for $f_s=0.15$, $0.22$, and $0.27$, respectively.
Here we use $E_J=400$~GHz.
}
\label{fig3}
\end{figure}

Below we
manipulate the superconducting AA around our example case $f=0.493$
(the two vertical dashed lines in Fig.~2) by changing the flux $\Phi_s$ threading
through the SQUID loop in three successive processes [see Fig.~1(b)]:

(i)~{\it Pumping process}:~First, the quantum circuit works at $f_s=0.22$.
We constantly pump the superconducting AA with an appropriate microwave field 
to make the state transition $|0\rangle\longrightarrow |2\rangle\equiv|E_2\rangle$
for a period of time.
Simultaneously, another microwave field is also used to trigger the transition
$|2\rangle\longrightarrow |1\rangle$. For $f_s=0.22$, $|t_{01}|\approx 0.01$,
$|t_{12}|\approx 0.07$, and $|t_{02}|\approx 0.13$. Because $|t_{01}|$ is about
one order of magnitude smaller than $|t_{12}|$ and $|t_{02}|$, a population inversion 
between the two lowest states $|0\rangle$ and $|1\rangle$ can be readily achieved.

(ii)~{\it Preserving the population inversion}:~We decrease $f_s$ to $f_s=0$.
Because $|t_{01}|$ now tends to zero, the transition
$|1\rangle\longrightarrow |0\rangle$ is forbidden. For typical values 
$E_J/h=400$~GHz and $\Phi_w^{(0)}/\Phi_0\sim 1\times 10^{-4}$ (see Sec.~V),
the state population inversion can be preserved for a time $\sim 0.02$~s, 
much longer than 
the photon lifetime $\tau_p\sim 8$~$\mu$s of a cavity with quality factor $Q=10^6$.

(iii)~{\it Switching-on process}:~We increase $f_s$ to $f_s=0.27$
to turn on the transition $|1\rangle\longrightarrow |0\rangle$
with an appreciable probability (i.e., $|t_{01}|\approx 0.13$, which is
one order of magnitude larger than $|t_{01}|\approx 0.01$ at $f_s=0.22$).

Here we emphasize that the two lowest energy levels are in {\it resonance} 
with the cavity 
mode when $f_s=0.27$. However, during the pumpimg, $f_s=0.22$. For this value of the 
reduced flux through the SQUID, the energy level difference between states 
$|0\rangle$ and $|1\rangle$ is appreciably different from that at $f_s=0.27$, so the 
two lowest levels at $f_s=0.22$ are very {\it off-resonance} with the cavity mode. 
Also, the level difference between $|0\rangle$ and $|2\rangle$ and that between 
$|1\rangle$ and $|2\rangle$ are very off-resonance with the cavity mode. 
Therefore, the qubit-cavity coupling is very weak during the pumping process, 
where we choose $f_s=0.22$ (instead of $f_s=0.27$, used for achieving strong 
qubit-cavity coupling).

\section{Micromaser and single-photon source}

\subsection{Interaction Hamiltonian}

Let us place the superconducting
AA in a quantum cavity, with the energy difference $E_1-E_0$ at $f=0.493$
and $f_s=0.27$ in resonance with the cavity mode.
In the switching-on process, the superconducting AA acts as
a two-level system and interacts with a single-mode quantized microwave field
via Rabi oscillations, i.e., a coherent exchange of energy between them.

In the subspace with basis states $|1\rangle$ and $|0\rangle$, the circulating
current can be written as 
\begin{equation}
I=\frac{1}{2}(A-B)\sigma_z + (C\sigma_{+} + \rm{H.c.}), 
\end{equation}
where $A=\langle 1|I|1\rangle$, $B=\langle 0|I|0\rangle$, $C=\langle 1|I|0\rangle$, 
and $\sigma_+=|1\rangle\langle 0|$ is the raising operator for the states of the 
two-level system. The quantized microwave field in a cavity can be written as 
\begin{equation}
\Phi_w =\Phi_w^{(0)} (a + a^{\dag}), 
\end{equation}
where $a$ ($a^{\dag}$) is the annihilation (creation) 
operator of photons of the cavity mode.
In the rotating-wave approximation, $(C\sigma_{+} + \rm{H.c.})\Phi_w$ becomes
$(t_{01}\sigma_{+}a + \rm{H.c.})$. 
Then, the interaction Hamiltonian (\ref{HA}) can be written as 
\begin{equation}
H'=-\frac{1}{2}(A-B)\Phi_w^{(0)}\sigma_z (a+a^{\dag})-(t_{01}\sigma_{+}a+\rm{H.c.}).
\label{HB}
\end{equation}
The first term on the right-hand side of Eq.~(\ref{HB}) only gives an effective 
contribution to the energy 
difference $E_1-E_0$ in the expression for the eigenvalues of the total 
Hamiltonian and it does not affect the Rabi oscillations. For the 
single-photon process we study, this contribution is a fixed value 
added to the energy difference $E_1-E_0$ and it can be included into the 
energy difference. Actually, this was explicitly shown for a charge 
qubit coupled via its SQUID loop to the cavity mode (see Ref.~\onlinecite{YN03}). 
Therefore, one can only consider the 
Jaynes-Cummimgs term for the interaction Hamiltonian: 
\begin{equation}
H'=-\hbar g(\sigma_+a + {\rm H.c.}),
\label{HC}
\end{equation}
where 
\begin{equation}
g=\frac{1}{\hbar}\,|t_{01}|.
\end{equation}
In Eq.~(\ref{HC}), we also ignore a phase factor that does not produce effects in 
our study. 

Below we estimate the contribution of the first term 
on the right-hand side of Eq.~(\ref{HB}). 
Usually, $A \sim -B \sim 0.5I_c$; 
in particular, $A=B=0$ at the degeneracy point $f=0.5$. Thus, we have 
\[
\frac{1}{2}(A-B)\Phi_w^{(0)} \sim 0.5I_c\Phi_w^{(0)}
=\pi E_J ({\Phi_w^{(0)}}/{\Phi_0}).
\] 
For a typical value of $\Phi_w^{(0)}/\Phi_0 \sim 10^{-4}$ in our case 
(see Sec.~V), 
\[
\frac{1}{2}(A-B)\Phi_w^{(0)} \sim 3\times 10^{-4}E_J,
\] 
which is much smaller than the energy difference 
$E_1-E_0 \sim 0.05E_J$ 
for $f=0.493$ and $f_s=0.27$. Thus, the first term on the right-hand side of  
Eq.~(\ref{HB}) can also be ignored here, even if its contribution is not included 
into the energy difference. 

\subsection{Photon statistics}

We assume that the interaction between the cavity and the AA is in the
strong-coupling regime, where the period $1/2g$ of the single-photon Rabi oscillations
is much shorter than both the relaxation time of the two-level system and
the average lifetime of the photon in the cavity.
After an interaction time $\tau$, the quantum circuit turns to the pumping and
population-inversion-preserving processes and it becomes ready for the next cycle of
the three successive processes described above.

If the superconducting AA is switched on at the times $t_i$ to interact
with the photons in the cavity, the time evolution of the density matrix $\rho$
of the cavity mode is governed by the map 
\begin{equation}
\rho(t_i + \tau)=M(\tau)\rho(t_i),
\end{equation}
where the gain operator $M(\tau)$ is defined as
\begin{eqnarray}
M(\tau)\rho\!&\!=\!&\!{\rm Tr}_a \left[\exp\left(-\frac{i}{\hbar}H'\tau\right)\rho
\right. 
\nonumber\\
&& \left. \otimes |1\rangle\langle 1| \exp\left(\frac{i}{\hbar}H'\tau/\hbar\right)
\right],
\end{eqnarray}
where ${\rm Tr}_a$ denotes the trace over the variables of the AA.
Here we {\it regularly} switch on the superconducting AA by periodically
repeating the cycle of the three successive processes described above.

With the cavity losses included, the dynamics of the density matrix $\rho$
is described by~\cite{ORZ}
\begin{equation}
\frac{d\rho}{dt}=r_a[M(\tau)-1]\rho - \frac{1}{2}r_a(M-1)^2\rho +
L\rho.
\label{density}
\end{equation}
In Eq.~(\ref{density}), $r_a$ is the switching-on rate for the superconducting AA.
The operator $L$ describes the dissipation of the cavity photon due to
a thermal bath:
\begin{eqnarray}
L\rho\!&\!=\!&\!-\,\frac{1}{2}\kappa (n_{\rm th}+1)(a^{\dag}a\rho+ \rho a^{\dag}a
-2a\rho a^{\dag}) \nonumber\\
&&\!-\,\frac{1}{2}\kappa n_{\rm th}(aa^{\dag}\rho + \rho aa^{\dag}-2a^{\dag}\rho a),
\end{eqnarray}
where $n_{\rm th}$ is the average number of thermal photons in the cavity and
$\kappa$ is the photon damping rate.

At steady state, $d\rho/dt=0$, which leads to a recursion relation for the steady
photon number distribution $p_n=\langle n|\rho|n\rangle$ of the cavity mode:
\begin{eqnarray}
&&p_{n+1}=\left\{\frac{n_{\rm th}}{n_{\rm th}+1}
+\frac{2N_t S(n+1)[1+\frac{1}{2}S(n+1)]}
{2(n_{\rm th}+1)(n+1)}\right\}p_n  \nonumber\\
&&~~~~~~~~~~ -\frac{N_t S(n+1)S(n)}
{2(n_{\rm th}+1)(n+1)}p_{n-1}\,,
\label{photon}
\end{eqnarray}
with initial condition
\begin{equation}
p_1=\left\{\frac{n_{\rm th}}{n_{\rm th}+1}
+\frac{2N_t S(1)[1+\frac{1}{2}S(1)]}
{2(n_{\rm th}+1)}\right\}p_0\,,
\end{equation}
where 
\begin{equation}
S(n)=\sin^2(g\tau\sqrt{n}),\;\;\; n=0,1,2,\dots. 
\end{equation} 
The quantity
\begin{equation}
N_t=\frac{r_a}{\kappa}
\end{equation}
represents the number of cycles for switching on the superconducting AA during the
photon lifetime of the cavity, and $p_0$ is determined by
$$\sum_{n=0}^{\infty}p_n=1.$$ 
Equation~(\ref{photon}) is very different from the recursion
relation for the atomic micromaser (see Ref.~\onlinecite{MASER}), 
where all the excited atoms enter the cavity at random times:
\begin{equation}
(n_{\rm th}+1)np_n=[n_{\rm th}n+N_t S(n)]p_{n-1},
\end{equation}
where 
\begin{equation}
N_t=\frac{{\overline r}_a}{\kappa}, 
\end{equation}
with ${\overline r}_a$ being the {\it average} injection rate of the excited atoms.

\begin{figure}
\includegraphics[width=3.4in,  
bbllx=42,bblly=392,bburx=548,bbury=740]{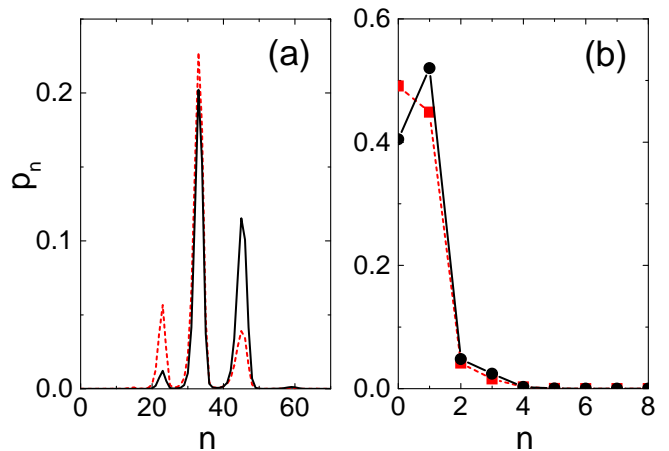} \caption{(Color
online) Steady-state photon statistics for $n_{\rm th}=0.1$, and
(a)~$N_t=100$, $\tau_{\rm int} (\equiv g\tau\sqrt{N_t})=10\pi$;
(b)~$N_t=1$, $\tau_{\rm int}=1.4\pi$.
These are results for a micromaser either using our quantum
circuit (black solid lines) or using natural atoms, e.g., rubidium (red
dashed lines). } 
\label{fig4}
\end{figure}

In Fig.~4(a), we present the steady-state photon statistics for
$N_t=100$. This statistics reveals an appreciable difference
between the micromaser with natural atoms and that with a
superconducting AA. Moreover, we show the steady-state photon
statistics for $N_t=1$ [see Fig.~4(b)]. It is striking that the
single-photon state has a probability at least one order of
magnitude {\it larger} than multi-photon states.  
Figure~4(b) shows that the photon statistics of the atomic micromaser looks 
similar to that of the new proposed micromaser, but the results of the atomic 
micromaser are derived by approximating the 
injection rate of the atoms with an average value ${\overline r}_a$. 
Indeed, the injection rate of the atoms into the micromaser 
has a distribution, instead of a fixed value. This is in sharp contrast to 
an AA micromaser having a fixed rate $r_a$ of switching on the AA. 

For the atomic micromaser, the excited atoms actually enter the cavity at random 
times and obey a Poissonian distribution. The number $n_{\rm ex}$ of
the excited atoms has a larger variance to the average number
$\overline{n}_{\rm ex}$:  
$\overline{\Delta n_{\rm ex}^2}=\overline{n}_{\rm ex}$, 
so larger fluctuations are expected for the
photon field in the cavity. In contrast, in the micromaser using a
SQC, the AA can be {\it regularly} switched on to interact with
the photons in the cavity and the photon-field fluctuations are
greatly reduced because $\overline{\Delta n_{\rm ex}^2}\sim 0$.
Therefore, one can use the SQC micromaser to implement a {\it
persistent} single-photon source with low-field fluctuations.

\section{Discussion}

\subsection{Experimentally accessible quantities}

Let us consider
a quasi-two-dimensional (2D) cavity, so that both the SQC and the cavity can be
fabricated on the same chip. Moreover, the SQC is placed at
an antinode of the cavity mode and the magnetic flux threads perpendicularly through
the SQC loop. The quantized magnetic flux inside the SQC loop can be
written as \cite{ORZ} 
\begin{equation}
\Phi_w=\Phi_w^{(0)}(a+a^{\dag}),
\end{equation}
with 
\begin{equation}
\Phi_w^{(0)}=\left(\frac{h\nu}{\epsilon_0 c^2Ah}\right)^{1/2}S_q,
\end{equation}
where $\nu$, $A$, $h$, and $S_q$ are the cavity frequency, the area of the quasi-2D
cavity, the thickness of the cavity, and the area of the SQC loop, respectively.
At $f=0.493$ and $f_s=0.27$, the numerical results in Fig.~2(c) give that
$E_1-E_0\approx 0.05E_J$.
When this level difference is in resonance with the cavity mode, the frequency of
the cavity mode is $\nu=(E_1-E_0)/h\approx 20$~GHz for a typical value of
$E_J/h=400$~GHz; the wavelength is $\lambda \approx 1.5$~cm.
Here, as an example, we use
$A\sim (1.5~{\rm cm})^2$ and $h\sim 1$~$\mu$m for the quasi-2D cavity.
Moreover, as shown in Ref.~\onlinecite{YNN}, the energy spectrum is nearly unchanged up to
$\beta_L\equiv L/L_J\sim 0.1$, where the Josephson inductance is defined by
$L_J= \Phi_0/2\pi I_c$. This gives a loop inductance $L\approx 40$~pH
and the diameter of the loop is about $32$~$\mu$m.
Then, we have 
\[
\Phi_w^{(0)}/\Phi_0\approx 1.1\times 10^{-4}.
\]
Also, at $f=0.493$ and $f_s=0.27$, the numerical results in Fig.~2(f) give that
$|t_{01}|=0.13I_c\Phi_w^{(0)}$. Using the values given above,
we obtain 
\[
g=\frac{1}{\hbar}|t_{01}|\approx 218~{\rm MHz}. 
\]
For $N_t=1$ and $g\tau\sqrt{N_t}=1.4\pi$ [cf.~Fig.4(b)],
the corresponding interaction time $\tau$ for the AA to couple with the cavity mode
in each cycle of operations is $\tau\approx 20$~ns.
This value of $\tau$ is experimentally feasible because
it is usually much shorter than the relaxation time $T_1$ of the flux qubit
and also can easily be much shorter than 
the photon lifetime $\tau_p$ of an experimentally accessible high-Q superconducting 
cavity (e.g., $\tau_p\sim 8$~$\mu$s for $Q=10^6$).

In Ref.~\onlinecite{CHIO}, a relaxation time of about $1~\mu$s was measured 
for a flux qubit, away from the degeneracy point.
For the JJs in that qubit, the ratio of the small to large junction is $\alpha=0.8$, 
which is close to $\alpha=0.77$ for Figs.~2(b) and 2(e) in this work. 
The longer relaxation time is due to a smaller transition matrix element 
$|t_{01}|$. In Fig. 2(e), $|t_{01}|$ is about $0.01$, when operating the flux qubit 
away from the degeneracy point (e.g., at $f=0.493$); in Ref.~\onlinecite{CHIO},
$\alpha=0.8$ and $|t_{01}|$ should be even smaller. 

In order to achieve a strong-coupling regime, we consider the 
case in Figs.~2(c) and 2(f), where $\alpha=0.66$ 
and $|t_{01}|$ becomes larger than $0.1$ at $f=0.493$. Now, $|t_{01}|$ 
increases by more than one order of magnitude. Thus, because the relaxation
time is proportional to $1/|t_{01}|^2$, the 
relaxation time should be shortened by two orders of magnitude, 
from about $1~\mu$s to about $10$~ns. In this case, the strong coupling 
between the flux qubit and cavity mode can be achieved, but the value 
of the relaxation time is comparable to the interaction time $20$~ns 
used for emitting a photon in the cavity.

Therefore, in our approach, we can use a novel way to change the parameter $\alpha$ 
by replacing the smaller junction with a tunable SQUID. Only when 
coupling the flux qubit with the cavity mode, we shift $\alpha$ to 
$\alpha=0.66$. When preparing the states of the flux qubit via 
realizing and preserving the state population inversion, we 
shift $\alpha$ to a smaller value, where the relaxation time can 
be long enough, even (in principle) arbitrarily long (in practice, 
it will not be infinitely long, but far longer than currently reachable). 
More importantly, we show that the process of changing 
$\alpha$ can be simultaneously adiabatic (so the state 
changes along the eigenstate) and fast enough (compared 
to the interaction time). This reveals its enhanced level of control 
and experimental feasibility.  

\subsection{Comparison with ordinary two-level system}

In our proposal, the circuit design not only can either enhance 
or reduce the relaxation 
time for the transition from $|1\rangle$ to $|0\rangle$, but can also switch on and 
off this transition by changing the externally applied magnetic flux 
inside the SQUID loop. In particular, a state population inversion 
can be created via the third level 
and preserved for a long time. Moreover,
when needed for generating a single photon, the SQC can be ``turned on", by changing 
an externally applied magnetic flux, to strongly couple to the cavity mode. 
These properties show that the circuit design provides an enhanced level of control,
as compared with ordinary two-level systems.

If one wants to just observe the fundamental vacuum Rabi oscillations, 
it is suitable to manipulate only the two lowest levels of the SQC. 
However, when applied to the micromaser proposed here, it is very disadvantageous 
to only access the two lowest levels, since it is hard to maintain 
a state population inversion with only these two levels.

If one uses the two lowest levels of the SQC, in order to strongly couple the SQC 
to the cavity mode, the resonance point should be near the degeneracy point 
of the SQC, where the transition between $|1\rangle$ and $|0\rangle$ is the strongest.  
As shown in Ref.~\onlinecite{NTT}, 
one can first operate the SQC far from the resonance point 
and prepare it in the excited state $|1\rangle$ by employing a $\pi$ pulse. 
Then, the $\pi$ pulse is followed by a shift pulse, which brings the system into 
resonance with the cavity mode. 
It is required that the process for shifting the qubit from 
the operating point to the resonance point is adiabatic, 
so as to still keep the qubit at the eigenstate $|1\rangle$ 
after this shifting process. Therefore,
due to the qubit-oscillation coupling, the excitation can be transferred
to the cavity, resulting in a single photon. However, this does not occur 
ideally.  As shown in Fig.~3(a), the quantity $K_{01}$ increases 
drastically when the reduced flux $f$ approaches the 
degeneracy point $f=0.5$. This means that the process 
for shiting the qubit from the operating point to the 
resonance point should be very slow (especially when 
approaching the resonance point) to keep it adiabatic. 
If so, a problem arises because of the strong relaxation 
from $|1\rangle$ to $|0\rangle$ around the resonance point, greatly 
reducing the probability of the single photon in the cavity.

Alternatively, one might consider using a nonadiabatic 
process to fast shift the flux qubit from the operating point to 
the resonance point. However, at the resonance point, 
the qubit state is not anymore the target eigenstate $|1\rangle$, 
but instead a superposition state of $|1\rangle$ and $|0\rangle$. This is 
problematic, also because it reduces the single-photon probability. 

However, as we have already shown, no such problems should occur in our proposal 
using three levels, 
because of the greatly enhanced level of control available when using 
three levels and controlling the $|0\rangle\longleftrightarrow |1\rangle$ 
transition strenght via the magnetic flux inside the SQUID loop. 

\subsection{Comparison with other theoretical works}

A theoretical work in Ref.~\onlinecite{ZHOU} considered a $\Lambda$-type three-level 
system, where {\it no} transition occurs between the lowest 
two levels, but transitions are allowed between the other levels. 
Together with the lowest two levels, the third level is used to 
indirectly achieve a one-qubit gate for the lowest two levels 
(i.e., the qubit levels). In that work, the circuit considered is 
an rf-SQUID consisting of a loop with one Josephson junction. It is known that 
this can be used as a flux qubit, but it requires a relatively large loop 
inductance to produce a two-well potential. This makes the 
rf-SQUID more susceptible to flux noise. 
Another theoretical work in Ref.~\onlinecite{AMIN} used the idea in 
Ref.~\onlinecite{ZHOU} to further show that 
a complete set of one-qubit gates could be achieved. 

However, in contrast to the $\Lambda$-type three-level system 
in Ref.~\onlinecite{ZHOU}, the lowest three levels of the circuit design in our 
approach can form a $\Delta$-type three-level system, when operating away from the 
degeneracy point. Most importantly, the transitions, particularly the transition 
between the lowest two levels can be {\it tuned} at will by simply changing 
the parameter $\alpha$ via the magnetic flux inside the 
SQUID loop. Therefore, this circuit can prepare and preserve 
the state population inversion between the lowest two levels, 
and can also trigger a strong transition between these two levels. 
Moreover, these processes provide an enhanced level of 
control and can be manipulated both adiabatically and fast enough. 
These remarkable properties do not exist in the system considered in 
Refs.~\onlinecite{ZHOU} and \onlinecite{AMIN}.

\section{Conclusion}

We have proposed a tunable on-chip micromaser using a SQC. 
The circuit design provides an enhanced level of control via adjusting 
the parameter $\alpha$ with the magnetic flux 
inside the SQUID loop. By taking advantage of the externally controllable
transitions between states, we can both produce and preserve a
state population inversion for the two working levels of the SQC.
When the previously generated photon in the cavity is decaying,
the SQC can generate a new single photon. These processes can be
regularly repeated to produce single photons in a persistent
manner. This approach provides a controllable way for implementing a
persistent single-photon source on a microelectronic chip.

\begin{acknowledgments}
This work was supported in part by the NSA and ARDA under AFOSR
contract No.~F49620-02-1-0334, and by the NSF grant No.~EIA-0130383.
J.Q.Y. was supported by the 
National Natural Science Foundation of China (NSFC)
grant Nos.~10474013, 10534060 and 10625416.
C.P.S. was partially supported by the NSFC and the NFRPC.
\end{acknowledgments}


\end{document}